\documentclass[twocolumn]{aastex61}
\pdfoutput=1 
\usepackage{amsmath,amstext}
\usepackage[T1]{fontenc}
\usepackage{apjfonts} 
\usepackage[figure,figure*]{hypcap}

\begin{document}

\title{Identification of the infrared counterpart of SGR\,$1935+2154$ with the {\em Hubble Space Telescope} }

\author{Andrew J. Levan}
\affiliation{Department of Physics, University of Warwick, Coventry, CV4 7AL, UK}

\author{Chryssa Kouveliotou}
\affiliation{Department of Physics, The George Washington University, Corcoran Hall, 725 21st St NW, Washington, DC 20052, USA}
\affiliation{GWU/Astronomy, Physics and Statistics Institute of Sciences (APSIS)}

\author{Andrew S. Fruchter} 
\affiliation{Space Telescope Science Institute, 3700 San Martin Drive, Baltimore, MD 21218, USA}

\email{A.J.Levan@warwick.ac.uk}

\keywords{stars: neutron --- supernovae: general}

\begin{abstract}
We present deep {\em Hubble Space Telescope} observations of a new magnetar source, the 
soft gamma-repeater SGR\,$1935+2154$, discovered by {\em Swift}. We obtained three epochs of observations: while the source was active in March 2015, during a quiescent period in August 2015, and during a further active phase in May 2016. Close to the center of the X-ray error region identified by {\em Chandra} we find a faint ($F140W(AB)=25.3$) source, which fades by a factor of $\sim2$ over the course of 5 months between the first two epochs of observations, before rebrightening during the second
active period. If this source is indeed the counterpart to SGR~$1935+2154$ then it is amongst the faintest yet located for a magnetar. 
Our observations are spaced over 1.3 years and enable us to place limits on the source velocity of $\mu = (60 \pm 40)$\,km s$^{-1}$ kpc$^{-1}$; observations on timescales of a decade can hence probe proper motion limits smaller than the velocities observed for the majority of pulsars. The comparison of the optical/IR and X-ray lightcurves of the source suggests that emission in the two regimes is associated but not directly correlated, offering support for a magnetospheric versus a fallback disc origin. 
\end{abstract}

\section{Introduction}
Magnetars are neutron stars with surface magnetic fields in excess of $10^{14}$ G \citep[e.g.,][]{kouveliotou98}; almost all sources emit short, soft X-ray bursts repeated in varying frequencies (from days to months to years). Although initially thought to be extremely rare, perhaps originating only from the most massive stars \citep[e.g.,][]{gaensler05}, increasing numbers of magnetars have been found in recent years from their burst activity
\citep[e.g.,][]{magcat,collazzi15}, thanks to sensitive wide-field X-ray/gamma-ray monitors such as the {\em Swift} Burst Alert Telescope (BAT) and the {\em Fermi}/Gamma-ray Burst Monitor (GBM). The total number of confirmed magnetars is now approaching 30. Interestingly, magnetar-like bursts have been detected from Rotation Powered Pulsars (RPPs), including PSR\,J$1846-0258$; \citep{gavriil08}, and PSR\,J$1119-6127$ \citep{camilo00,gogus16}, possibly providing the link between the two populations. This link was strengthened with the detection of magnetar sources with low magnetic fields well within the realm of RPPs, such as SGR\,$0418+5729$ ($B = 6\times10^{12}$ G; \cite{rea10}, Swift\,J$1822.3−1606$ ($B = 1\times10^{13}$ G; \cite{scholz14}) and 3XMM\,J$185246.6+003317$ ($B<4\times10^{13}$ G; \cite{rea14}). It appears, therefore, that there may be a continuum between high $B-$field RPPs and magnetars, indicating a common origin and evolution between the two populations. 

Most magnetars appear to be young (characteristic spin-down ages $t_c = P / 2\dot{P} \sim 10^{3}-10^{4}$ years) suggesting they can only be detected for a short time after their formation. Long quiescent intervals between periods of strong bursts make obtaining a complete census of young magnetars in the Milky Way very difficult, if not impossible. Estimates of the fraction of the SNe resulting in magnetars versus ``normal'' pulsars have been made in the past with population synthesis models \citep{pop06} and with simple calculations based on their ages and detected source numbers \citep{kouv94}; both concluded that approximately 10\% of the neutron stars formed in SNe could be magnetars.

However, combining typical magnetar lifetimes ($\sim 10^4$ years) with the Galactic core collapse SN rate ranging between 1-3 per century \citep{diehl06} would require that between $\sim30\% -10\%$ of the Milky Way SNe have created the currently known magnetar population. Most likely the rate is even higher, as many more remain to be found. While this is a very crude estimate, such rates indicate that magnetar progenitors must be drawn from much further down the stellar mass function in order to explain the observed rate, something that has observational support in some cases \citep[e.g.,][]{davies09}. Understanding which stars create magnetars, either from detailed studies of magnetars themselves or from their environments, can, therefore, provide unique constraints on the diversity of the final stages of stellar evolution.

Magnetars are increasingly invoked to explain  many of the most energetic events in nature, including both long- and short-duration gamma-ray bursts and super luminous supernovae
\citep[e.g.,][]{metzger15,rowlinson13}. While it is apparent that these events are far too rare to be linked to the $\sim30$ young magnetars observed in the Milky Way \citep[see e.g.,][]{rea15}, recent work has suggested that magnetars may actually have played an important role as engines shaping many SNe \citep{sukhbold17,piran17}, and hence understanding the frequency of their production, their progenitors, and in some cases their associated supernova remnants is of significant interest.

Magnetar discoveries are extremely challenging. Most sources have been detected when they enter burst active outbursts, emitting very short ($\sim$100~ms), soft X-ray bursts at varying intervals and numbers (from several hundred per day to less than 10 per year). Bursting activity is almost always associated with an increase in X-ray luminosity that may last from days to months.  While X-ray counterparts have been readily identified in the majority of cases, multi-wavelength observations are needed to accurately pinpoint the source and constrain the origins of the
emission and the source dynamics via proper motion \citep[e.g.,][]{tendulkar12,tendulkar13}. Unfortunately, due to the small scale heights of Galactic magnetars ($\sim20$\,pc) such observations are often plagued by crowded fields and significant line of sight extinction, besides faint counterparts. Indeed, while optical counterparts have now been found to a number of sources \citep[see e.g.,][for a review of optical/IR observations]{magcat,rea11}, these are exclusively faint, and the majority are hard to study without adaptive optics, or with the {\em Hubble Space Telescope (HST)}, an approach we use here for observations of SGR\,$ 1935+2154$.

\begin{figure*}[ht]
    \centering
    \includegraphics[width=16cm,angle=0]{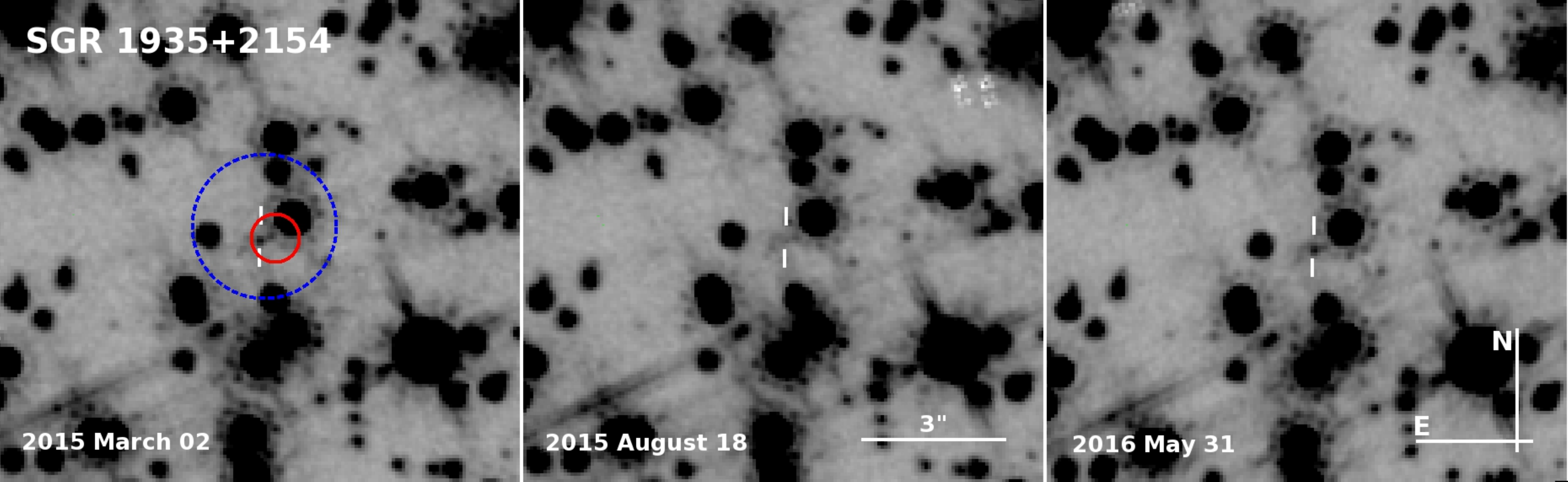}
\caption{Our {\em Hubble Space Telescope} observations of SGR\,$ 1935+2154$. The {\em Chandra} location is marked with a red circle and the {\em Swift} enhanced position is shown by the larger blue dashed circle (the white regions in the NE of the second epoch are known detector defects). Close to the center of the error box is an apparently transient source, fading by $\sim 0.8$ magnitudes between the two 
observations. It is the only transient of this significance within $\sim 20$\arcsec\, of this location, and we identify it, therefore, as the counterpart of SGR\,$1935+2154$.}
\label{finder}
\end{figure*}

\section{Observations}

\subsection{Discovery of SGR\,$1935+2154$}
SGR\,$1935+2154$ triggered the {\em Swift}/Burst Alert Telescope on 2014 July 5 \citep{lien14a}, when it emitted a short, magnetar-like burst.
Follow up observations with the {\em Swift}/X-ray Telescope (XRT) identified an X-ray
counterpart; subsequent observations with {\em Chandra} enabled measurement of the source period (P=3.24729 s) \citep{israel14}. Finally, observations with {\em NuSTAR} and {\em XMM-Newton} confirmed the period and provided the period derivative 
($\dot{P} = 1.43 \times 10^{-11}$ s s$^{-1}$), from which a dipole field of $B=7 \times 10^{14}$G and a characteristic age of $\tau_c = P / 2\dot{P} \sim3600$ years were calculated \citep{younes15,israel16,younes17}. Following this initial detection a further period of activity was detected with GBM, beginning on 2015 February 22 \citep{burns15}, suggesting the source had entered a new active phase. The later stages of this phase included a rare so-called intermediate burst (lying between the frequent faint bursts seen in many magnetars and the very rare Giant flares \citep{kouv01} on 2015 April 12 \citep{kozlova16}. After a short period of quiescence the source became active again in May 2016 \citep{younes16,kozlova16b}. Detailed $\gamma$-ray and X-ray observations of SGR\,$1935+2154$ are presented in \cite{israel16} and \cite{younes17}.

\subsection{Hubble Space Telescope}

We obtained three epochs of observations of SGR\,$1935+2154$ with {\em HST}. Observations consisted of a single orbit with the F140W filter to maximize throughput. The orients were set at 90$^{\circ}$ or 180$^{\circ}$ relative to previous epochs so that diffraction spikes from bright stars overlaid each other. At each epoch, we took 4x600~s of observations. Individual exposures were aligned utilizing point sources in common to each image, via the {\tt tweakreg} task; finally the images were drizzled \citep{fruchter02} onto a common frame with an output scale of 0.07\arcsec. The RMS of {\tt tweakreg} was tested by a direct map between point sources
 in the drizzled images and found to be $\sim 8$ mas, setting the level of relative astrometry that can be obtained. Finally, we refined the world coordinate system using $\sim 100$ 2MASS stars, providing an absolute position accurate to better than 0.2\arcsec.

\section{The infrared counterpart of SGR\,$1935+2154$}
The field of SGR\,$1935+2154$ was observed with {\em Chandra}/ACIS-S on 2014 July 15 for a duration of 10 ks. The source position was determined to be at  RA=19:34:55.5978, DEC=21:53:47.7864, with an accuracy of 0.6\arcsec  (90\% confidence; \citet{israel16}).

Our {\em HST} observations reveal one source within the {\em Chandra} error circle, with several more close to, but formally outside this region. The source within the error circle was
clearly variable between our observations, strengthening its identification as the IR magnetar counterpart. No other variables were detected close to the X-ray location based on image subtraction, and the photometry of nearby stars was consistent between the images. 
Because of the crowded field, we performed photometry in a $0.1\arcsec$ radius aperture, and used published aperture corrections. Background estimation was achieved by placing apertures of the same size through the image away from obvious bright sources, although the field is sufficiently crowded that some background from faint stellar or diffuse objects may have been included. The resulting photometry is shown in Table~\ref{phot}.

The IR counterpart location is at RA=19:34:55.606 DEC=21:53:47.45 ($\pm 0.2$\arcsec). We note that the photometry of the source in progressively larger apertures provides a brighter source even after correcting for the encircled energy, suggesting that there is a diffuse component, and implies that, even with the resolution of {\em HST}, precise photometry may be difficult. However, we also note that the measured fading is not strongly dependent on the aperture size, for aperture sizes $<0.3$\arcsec; beyond this, the increased background contribution and contamination from nearby stars begins to be significant. 

There is no measured proper motion of the source between the first and final images taken 1.3 years apart, with the formal motion measured as $\mu = 13 \pm 9$\,mas yr$^{-1}$, including the contributions from the astrometric tie ($\sim$8\,mas) and centroiding ($\sim$4\,mas). This assumes the source is point-like and may be subject to some additional uncertainty if underlying sources contributions could be accounted for. This issue could be resolved with future long-term observations with difference imaging.

\begin{table}
\caption{{\em HST} observations of SGR\,$1935+2154$}
\begin{center}
\begin{tabular}{llll}
\hline
Date & MJD & AB-mag \\
\hline
2015-03-02:23:04:06 & 57083.96118 &  25.26 $\pm$ 0.04\\
2015-08-18:08:11:06 & 57252.34104 &  26.09 $\pm$ 0.09\\
2016-05-31:22:13:51 & 57555.92685 &  25.22 $\pm$ 0.05\\
\hline
\end{tabular}
\end{center}
\tablecomments{Photometry has been obtained in 0\farcs1 apertures centered on the location of the counterpart and corrected for encircled energy.}
\label{phot}
\end{table}%

\section{Discussion}
Numerous searches have been conducted for the optical counterparts of magnetars. To date there are optical/IR detections of $\sim11$ out of $29$ magnetars\footnote{based on the McGill Magnetar database, and references therein. These numbers are, however, approximate as there are some cases in which plausible candidates have been observed, based on X-ray locations, or unusual colours, but for which a secure identification has yet to be made.}. Most of these detections have been made in ground-based K-band observations; the detected magnitudes are in the range $18-22$ \citep[e.g.,][]{magcat}. The counterpart of SGR\,$1935+2154$ has a discovery magnitude of F140W(AB) $\sim 25.3$ (H(Vega)=24.0), making it  amongst the faintest discovered to date. Only the recent identification of 2E 1613.5-5053, the central compact object of the SN remnant RCW 103, appears fainter \citep{tendulkar17}, although is also in a bluer band, where foreground extinction is a larger concern, and is somewhat brighter in the slightly redder F160W observations. However, there are {\em HST} observations that place deeper limits on a handful of magnetars - most notably observations of SGR~$0418+5729$ \citep{durant11}, which was located in a region with relatively little crowding, where deep J-band observations place a limit of F110W$<$27.4. This depth is significantly fainter than our detection of SGR\,$1935+2154$, despite the likely location of SGR~$0418+5729$ in the Perseus arm at a distance of only $\sim 2$kpc, which is closer than the likely distance of SGR~$1935+2154$ (see section 4). Interestingly, SGR~$0418+5729$ has a very slow spin-down rate implying an unusually low field \citep{rea10} and a large age. It is not clear whether the non-detection of an IR counterpart for this source depends on its large age or the fact that the IR observations were taken long after its active period ended. 

\begin{figure*}
    \centering
    \includegraphics[width=14cm,angle=0]{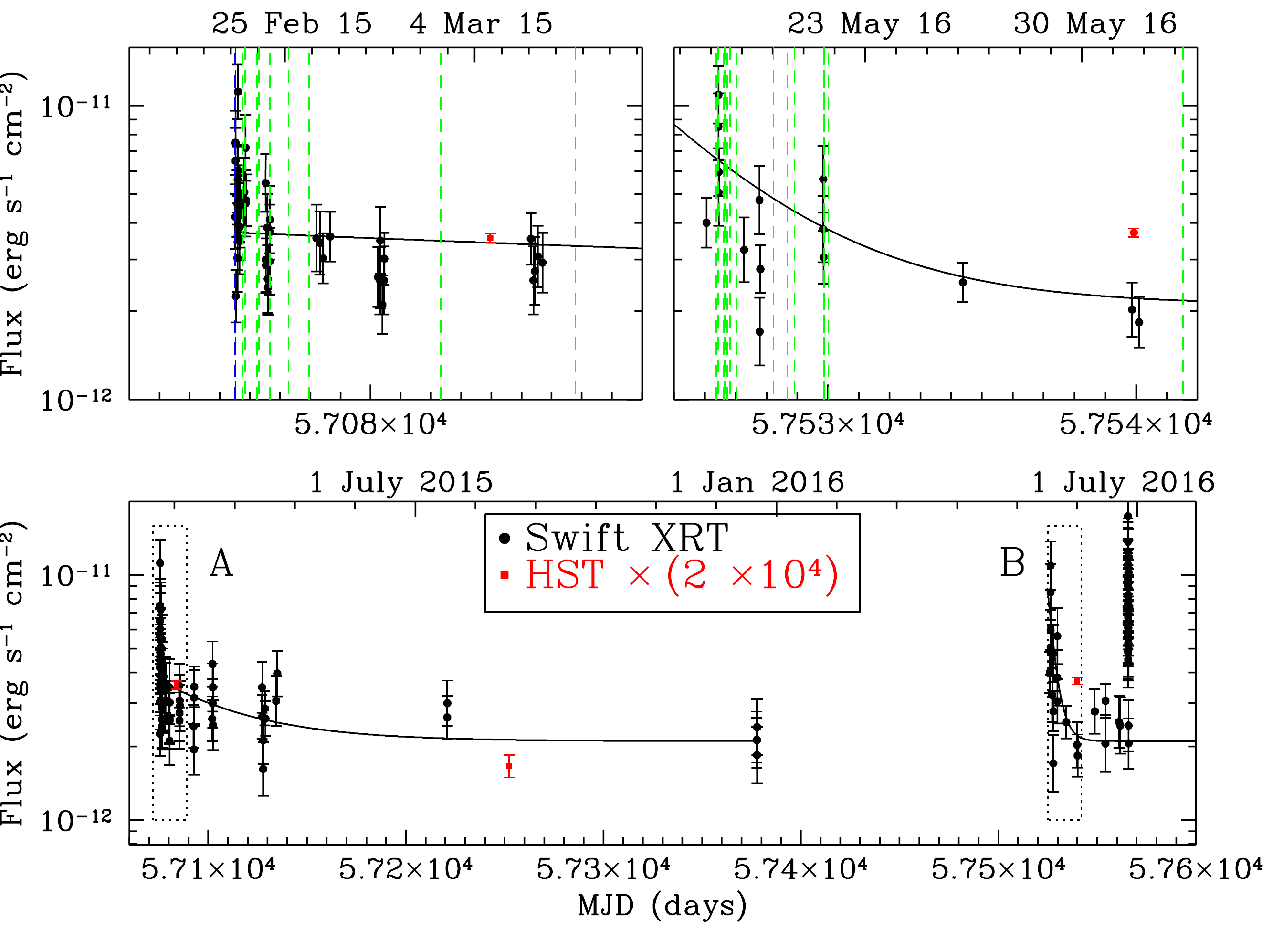}
\caption{The X-ray (black) and infrared (red) light curve of SGR~$1935+2154$. The lower panel shows a long (1.5 year) time range, while the top panels are zoomed in around the periods of enhanced activity. In these panels the dashed vertical lines show the times of the bursts from {\em Swift}/BAT (blue) and {\em Fermi}/GBM (green); the black line is the fit to each outburst from \cite{younes17}. The {\em HST} observations were obtained during the active periods, but not during their peaks.}
\label{irlc}
\end{figure*}

In Figure~\ref{irlc} we compare the X-ray and IR light curves of SGR\,$1935+2154$. The X-ray data shown are from the {\em Swift}/XRT (utilizing the methods of \cite{evans09}) in the $0.3-10$ keV range with an assumed counts to flux conversion of $10^{-10}$ erg s$^{-1}$ cm$^{-2}$ count$^{-1}$, suitable for the spectrum observed (we assume no X-ray spectral variability; \cite{israel16,younes17}) and use an approximate value). This is compared to the $\nu F_{\nu}$ IR observations. The X-ray flux evolution has been modeled as an exponential decay \citep{younes17}, but with very different time constants for the 2015 and 2016 outbursts of 43 and 3.7 days. These models are also plotted in Figure~\ref{irlc}.

While we do not have archival IR observations to establish a quiescent level well away from any activity, it is clear that at least in the case of the February 2016 outburst there was increased IR emission. However, this does not appear to directly track the X-ray observations. In particular,  for the May 2016 outburst, the X-ray emission returns rapidly to quiescence on a timescale of a few days \citep{younes17}, a decay time a factor $\sim 10$ shorter than in the February 2016 activation. However, the IR observations reveal a source at a similar brightness to the earlier outburst. This means that the X-ray to IR ratio is different by a factor of $\sim 2-3$ between the first (March 2016) and final (May 2017) set of {\em HST} observations (see Fig 2). 
This suggests that the origin of the IR emission is associated but not directly correlated to the X-rays.

The X-ray spectra can be fit with a combination of a blackbody (BB) and a power-law (PL) component \citep{israel16,younes17}. The BB has a temperature of $\sim0.5$ keV ($\sim 5 \times 10^6$~K), peaking at 2 keV or 6\AA. Extrapolating this BB to the IR predicts a magnitude $>37$, far too faint for detection. Moreover, the BB component does not seem to be variable during outbursts \citep{younes17}, and thus would not naturally explain the IR variation. However, the harder component does appear to vary during outbursts and its extension to IR is a more plausible explanation. {\em XMM-Newton}, {\em Chandra}, and {\em Swift} observations suggest that the PL has a photon index of $\Gamma \sim 2$, in which case extrapolation to the IR predicts F140W(AB) magnitudes around 22, comfortably explaining the IR counterpart emission (even accounting for extinction). However, the harder PL emission indicated by observations from {\em NuSTAR}, with $\Gamma=0.9$ would predict a source $\sim 100$ times fainter than observed. Clearly, the uncertainty of these results does not warrant extrapolating fluxes over a factor $>1000$ in frequency. 

Alternatively, the IR emission could arise from a fallback disc, as has been suggested for other magnetars \citep{wang06,kaplan09}; indeed models have been developed in which passive illumination of a fallback disc can drive correlated X-ray and IR emission in magnetar activations (see e.g., \cite{ertan}). However, the directly correlated emission is apparently rare (see section 4.1) and this argues against such an interpretation and points towards a magnetospheric origin both for the enhanced X-ray and the IR flux \citep{tam04,wang08}. 

The existence of such discs should be readily testable with observations in the near to mid-IR. 
Mid-IR imaging of the field has been taken as part of the {\em WISE} survey from 3.4-22 microns, and with the {\em Spitzer} GLIMPSE project from 3.6-24 microns. However, the poor point spread function of these surveys makes confusion a major issue at low Galactic lattitudes, and the limiting magnitudes of the surveys are such that we would not expect to detect the counterpart even with extreme colors. For example, even without confusion, detection would require a factor $>100$ brightness increase from F140W to 3.6 microns given the typical WISE 5$\sigma$ limiting magnitude of $\sim 19.3$ (AB)\footnote{http://wise2.ipac.caltech.edu/docs/release/allsky/}. Given the confusion issues, and shallow depth we do not attempt to extract formal limits from the data; we note, however, that there is no extended mid-IR emission as seen in the case of SGR\,1900+14 \citep{wachter08}.
With the advent of {\em JWST} it will be possible for the first time to obtain high resolution mid-IR observations particularly sensitive to the spectral signature of any fallback disc emission around magnetars.

\subsection{Comparison with other magnetar counterparts}
\begin{figure}[ht]
    \centering
    \includegraphics[width=7.5cm,angle=0]{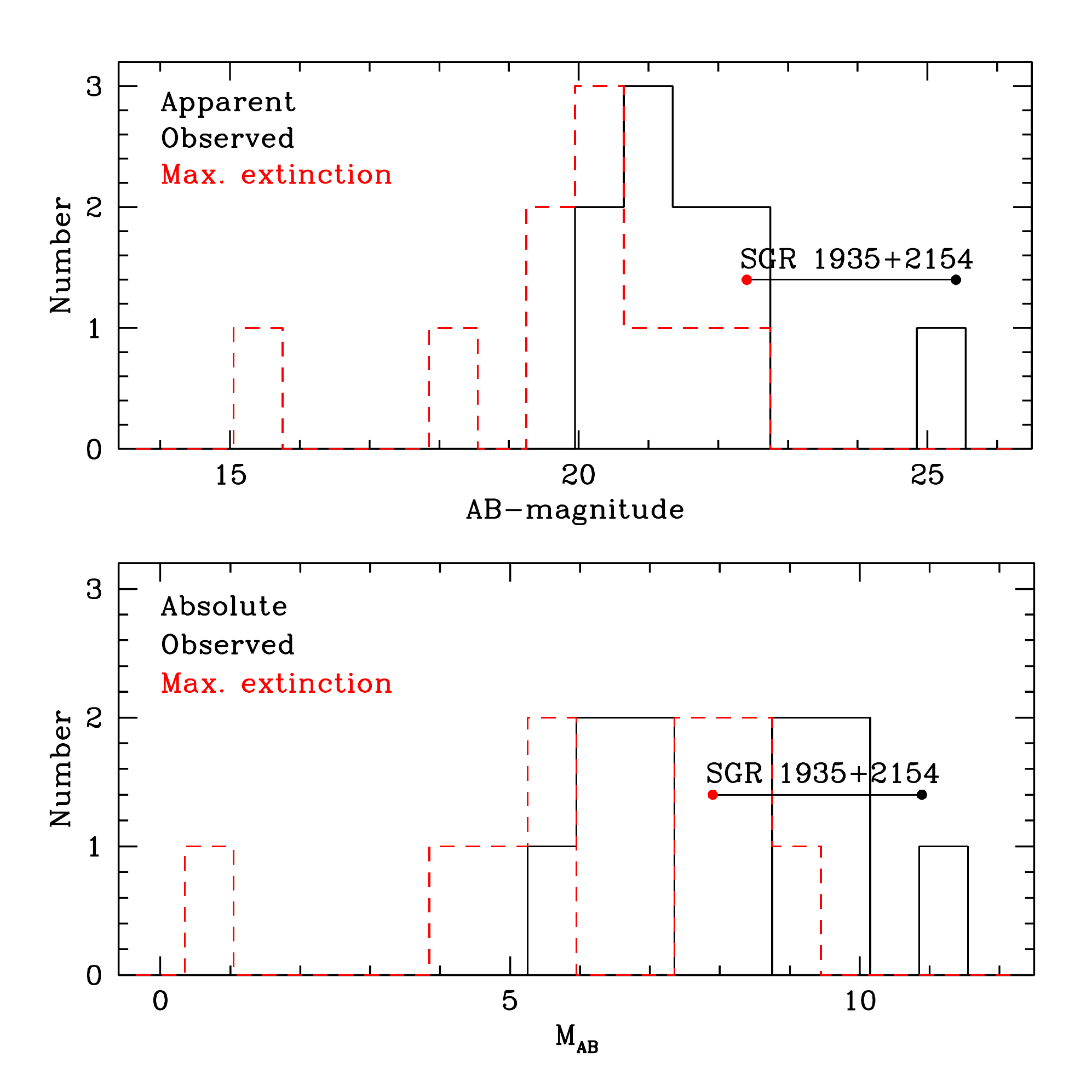}
\caption{The comparative properties of the optical/IR counterparts of magnetars \citep[from][]{magcat}. {\bf Top panel:} The distribution of apparent magnitudes of the brightest detections (solid line) corrected for maximal extinction, defined as the entire Galactic extinction as measured by \cite{schlafly}. {\bf Bottom panel:} As for the top panel, but plotted for absolute magnitudes, based on the best available estimates for the distances to magnetars \citep[again from][]{magcat}. The range for SGR\,$1935+2154$ is shown in each panel as a solid horizontal line. While there is significant uncertainty in determining the physical parameters of the optical/IR counterparts, it is apparent that, while observationally faint, the counterpart of SGR\,$1935+2154$ is broadly consistent with the properties of other counterparts (here we assumed a distance of 8\,kpc for SGR\,$1935+2154$). Note that other magnetars are based on ground-based K-band observations, so that a colour term needs to be applied between the F140W and these. Note: To avoid having to consider multiple color terms (in different directions) we only compare to ground based K-band observations. }
\label{ircomp}
\end{figure}

We compare the properties of the counterpart of SGR\,$1935+2154$ with those of other magnetars by plotting the brightest optical/IR detections for each source in Figure~\ref{ircomp}. In doing so, we show both the observed distribution of magnitudes and the same distributions taking into account uncertainties relating to foreground extinction and likely source distances. We find that  while the counterpart to SGR~$1935+2154$ is  particularly faint, it is not distinct from those of other magnetar counterparts, and is not necessarily of especially low luminosity. The quiescent and outburst X-ray fluxes for SGR\,$1935+2154$ are relatively faint compared to other magnetars \citep{magcat}, while the X-ray to IR flux ratio (in $\nu F_{\nu}$ before absorption/extinction correction) is $\sim 5000$. This suggests that for similar flux ratios {\em HST} observations will be a powerful route to identifying counterparts for much of the Galactic magnetar population. In the near future with {\em JWST} affording the combination of larger aperture and redder filter system (less affected by foreground extinction) we should be able to make snapshot colour identification of the counterparts of the majority of magnetars relatively routine. 

In addition to the bulk photometric properties of the magnetar candidates, we can also compare the evolution in tandem of their optical/IR and X-ray emission following outbursts, which could be highly diagnostic of emission mechanisms. However, the limited observations of magnetar optical counterparts to date do not as yet provide a clear picture. Thus far, only magnetar 1E~$2259+586$ shows a clear X-ray to optical/IR correlation, with
identical decay indices following an outburst\citep{tam04}. Other observations of different sources provide less clear and at times contradictory results. There is, e.g., a claim of correlated emission for XTE\,J1810-197 \cite{rea04}, based on late time X-ray and optical observations, but other analyses of the same magnetar with a larger sample of IR observations from the VLT \citep{camilo07} suggest significant optical/IR variability, compared to a monotonic X-ray decay.

\cite{tam08} have reported on the results of multi-year observations of 1E\,1048.1-5937, which indicate a complex behaviour even within this single source. For example, in 2002 the IR and X-ray emissions appeared to be anti-correlated, while in 2007 there was a significant rise in the IR flux at the time of an X-ray flare. Even the well studied case of 4U\,0142+61 does not show a clear correlation between the persistent and the pulsed fluxes of the X-ray and IR/optical wavelengths \citep{durant06}. Finally, \cite{tendulkar17} observed 2E\,$1613.5-5053$ with HST shortly after it a magnetar-like outburst was detected by {\em Swift} \citep{dai16,rea16}. They report an IR brightening coincident with the rise in X-rays, albeit at lesser brightness, e.g., at a factor of $\sim 3$, compared to a factor of 7 in the X-rays.

It is plausible that this variety of behaviour suggests that there are a range of physical processes driving the X-ray and IR emission, and that the relative contribution of these in a given source is responsible for the diversity in emission. Indeed, the apparent difference in the outburst amplitudes could readily be explained by different relative X-ray and optical/IR levels in quiescence, something that seems likely. Alternatively, it may be that our understanding of the activity is incomplete, for example due to outbursts being missed, or because observations are not truly simultaneous in many cases (something that may be important close to outburst when the source may be evolving rapidly). Future observations to resolve this question require the intensive combination of cadence, depth (since the sources are faint) and co-ordination between different observatories.

\subsection{The broader environment and birthplace}

SGR\,$1935+2154$ is located close to the Galactic plane at $l=57.25$, $b=0.82$. This direction lies along the Orion Spur (where the Sun lies) and crosses the Perseus Arm at $8-9$\,kpc distance. The significant overdensities of stars in this direction lie in these two regions, with most stars in the Perseus Arm. X-ray observations measure a $N_H$ in this direction higher than the total Galactic column (N$_H$=(1.6 $\pm$ 0.2 )x10$^{22}$  vs $N_H$(gal) = 1.23 $\times 10^{22}$ cm$^{-2}$, \cite{israel16,willingale}), suggesting the source is more likely at a larger distance, and may have some additional absorption. The IR extinction in this direction is E(B-V) = 4.31 \citep{schlafly}, corresponding to $A_{F140W} = 2.99$.

It has been suggested that SGR~$1935+2154$ is associated with the supernova remnant G\,$57.2+0.8$ \citep{gaensler14}, in which the magnetar is almost central. The SNR is approximately 0.1$^{\circ}$ in radius, corresponding to $\sim 20$pc at a favoured distance of $11.7\pm 2.8$\,kpc \citep{surnis16} somewhat beyond the bulk of the Perseus Arm. Its size is comparable to young ($<20$-kyr) SNRs in the LMC \citep{badenes10}. We note that the size of the remnant would be inconsistent with the characteristic age of the magnetar were it located significantly closer (e.g., at 1\,kpc within the Orion Spur); this offers further credence to the larger distance estimate. 

Our measured motion of $\mu =13 \pm 9$\,mas corresponds to a physical velocity of ($\mu \sim 60 \pm 40$)\,km s$^{-1}$ kpc$^{-1}$. For a distance of 10\,kpc our 2$\sigma$ upper limit would be in the region of 800\,km s$^{-1}$, something achieved by only the fastest $\sim 10-20$\% of neutron stars \citep{arzoumanian,hobbs05}. The location of the SGR within the SNR is probably within $\sim 30\arcsec$ of the centre, given the difficulties of determining the centre in existing radio imaging. This constrains the velocity to the $<400$ km s$^{-1}$, comparable to the proper motion constraints. Given our intrinsic astrometric accuracy, observations on a timescale of $5-10$ years would place proper motion limits lower than those of a typical neutron star ($80-160$\,km s$^{-1}$).

There is no obvious sign of an underlying cluster environment as might be expected for a massive progenitor; sources close to the magnetar location have similar colours to those at larger radii. The region lies in the field of view of the IPHAS survey \citep{drew05}, but no H$\alpha$ emission is seen near the magnetar location, which is not surprising given the high foreground extinction. Finally, there is some evidence for diffuse emission underlying the source position in our {\em HST} imaging, but given the high crowding of the field it is difficult to ascertain if this is true diffuse emission, or due to one or more fainter point-like sources.

\section{Summary}
The optical/IR counterpart of SGR\,$1935+2154$ is  one of the faintest yet detected. Given the very crowded nature of the field it is clear that deep observations of X-ray locations will seldom allow for the unique identification of counterparts \citep[e.g.,][]{durant05,durant08}, in the absence of information about either colour or variability. The extremely faint counterpart of the magnetar would have been extremely challenging for ground-based observatories, requiring several nights of integration with ground-based AO for detection. For many magnetars, therefore, the best chance of counterpart detection and identification with current technology may well be rapid response observations with {\em HST}. Such observations would ultimately enable light curve monitoring, the construction of spectral energy distributions and the measurements of proper motions. 

\section*{Acknowledgements}
We thank the referee for a constructive and timely report that improved the quality of the paper. We thank the staff of STScI for the rapid execution of these observations, and the {\em Swift} team for their scheduling work. This project has received funding from the European Research Council (ERC) under the European Union's Horizon 2020 research and innovation programme (grant agreement no 725246). We acknowledge support from the UK Science and Technology Facilities Council 
under grant number ST/L000733/1 and ST/P000495/1. Based on observations made with the NASA/ESA Hubble Space Telescope, obtained at the Space Telescope Science Institute, which is operated by the Association of Universities for Research in Astronomy, Inc., under NASA contract NAS 5-26555. These observations are associated with programs 14055 and 14502.


\begin{thebibliography}{99}
\expandafter\ifx\csname natexlab\endcsname\relax\def\natexlab#1{#1}\fi
\bibitem[Arzoumanian et al.(2002)]{arzoumanian} Arzoumanian, Z., Chernoff, D.~F., \& Cordes, J.~M.\ 2002, \apj, 568, 289 


\bibitem[{{Badenes} {et~al.}(2010){Badenes}, {Maoz}, \& {Draine}}]{badenes10}
{Badenes}, C., {Maoz}, D., \& {Draine}, B.~T. 2010, \mnras, 407, 1301

\bibitem[{{Burns} \& {Younes}(2015)}]{burns15}
{Burns}, E., \& {Younes}, G. 2015, GRB Coordinates Network, 17496

\bibitem[Camilo et al.(2000)]{camilo00} Camilo, F., Kaspi, V.~M., Lyne, A.~G., et al.\ 2000, \apj, 541, 367 


\bibitem[Camilo et al.(2007)]{camilo07} Camilo, F., Ransom, S.~M., Pe{\~n}alver, J., et al.\ 2007, \apj, 669, 561 


\bibitem[Collazzi et al.(2015)]{collazzi15} Collazzi, A.~C., Kouveliotou, C., van der Horst, A.~J., et al.\ 2015, \apjs, 218, 11 



\bibitem[D'A{\`i} et al.(2016)]{dai16} D'A{\`i}, A., Evans, P.~A., Burrows, D.~N., et al.\ 2016, \mnras, 463, 2394 



\bibitem[{{Davies} {et~al.}(2009){Davies}, {Figer}, {Kudritzki}, {Trombley},
  {Kouveliotou}, \& {Wachter}}]{davies09}
{Davies}, B., {Figer}, D.~F., {Kudritzki}, R.-P., {et~al.} 2009, \apj, 707, 844



\bibitem[Diehl et al.(2006)]{diehl06} Diehl, R., Halloin, H., Kretschmer, K., et al.\ 2006, \nat, 439, 45 

\bibitem[{{Durant}(2005)}]{durant05}{Durant}, M. 2005, \apj, 632, 563

\bibitem[Durant \& van Kerkwijk(2008)]{durant08} Durant, M., \& van Kerkwijk, M.~H.\ 2008, \apj, 680, 1394-1397 


\bibitem[{{Durant} {et~al.}(2011){Durant}, {Kargaltsev}, \&
  {Pavlov}}]{durant11}
{Durant}, M., {Kargaltsev}, O., \& {Pavlov}, G.~G. 2011, \apj, 742, 77

\bibitem[Drew et al.(2005)]{drew05} Drew, J.~E., Greimel, R., Irwin, M.~J., et al.\ 2005, \mnras, 362, 753 


\bibitem[Durant \& van Kerkwijk(2006)]{durant06} Durant, M., \& van Kerkwijk, M.~H.\ 2006, \apj, 652, 576 


\bibitem[Ertan et al.(2006)]{ertan} Ertan, {\"U}., G{\"o}{\v g}{\"u}{\c s}, E., \& Alpar, M.~A.\ 2006, \apj, 640, 435 


\bibitem[Evans et al.(2009)]{evans09} Evans, P.~A., Beardmore, A.~P., Page, K.~L., et al.\ 2009, \mnras, 397, 1177 


\bibitem[{{Fruchter} \& {Hook}(2002)}]{fruchter02}
{Fruchter}, A.~S., \& {Hook}, R.~N. 2002, \pasp, 114, 144

\bibitem[{{Gaensler}(2014)}]{gaensler14}
{Gaensler}, B.~M. 2014, GRB Coordinates Network, 16533

\bibitem[{{Gaensler} {et~al.}(2005){Gaensler}, {McClure-Griffiths}, {Oey},
  {Haverkorn}, {Dickey}, \& {Green}}]{gaensler05}
{Gaensler}, B.~M., {McClure-Griffiths}, N.~M., {Oey}, M.~S., {et~al.} 2005,
  \apjl, 620, L95

\bibitem[Gavriil et al.(2008)]{gavriil08} Gavriil, F.~P., Gonzalez, M.~E., Gotthelf, E.~V., et al.\ 2008, Science, 319, 1802 


\bibitem[G{\"o}{\u g}{\"u}{\c s} et al.(2016)]{gogus16} G{\"o}{\u g}{\"u}{\c s}, E., Lin, L., Kaneko, Y., et al.\ 2016, \apjl, 829, L25 


\bibitem[Hobbs et al.(2005)]{hobbs05} Hobbs, G., Lorimer, D.~R., Lyne, A.~G., \& Kramer, M.\ 2005, \mnras, 360, 974 



\bibitem[Israel et al.(2005)]{israel05} Israel, G., Covino, S., Mignani, R., et al.\ 2005, \aap, 438, L1 


\bibitem[{{Israel} {et~al.}(2014){Israel}, {Rea}, {Zelati}, {Esposito},
  {Burgay}, {Mereghetti}, {Possenti}, \& {Tiengo}}]{israel14}
{Israel}, G.~L., {Rea}, N., {Zelati}, F.~C., {et~al.} 2014, The Astronomer's
  Telegram, 6370

\bibitem[{{Israel} {et~al.}(2016){Israel}, {Esposito}, {Rea}, {Coti Zelati},
  {Tiengo}, {Campana}, {Mereghetti}, {Rodriguez Castillo}, {G{\"o}tz},
  {Burgay}, {Possenti}, {Zane}, {Turolla}, {Perna}, {Cannizzaro}, \&
  {Pons}}]{israel16}
{Israel}, G.~L., {Esposito}, P., {Rea}, N., {et~al.} 2016, \mnras, 457, 3448

\bibitem[Kaplan et al.(2009)]{kaplan09} Kaplan, D.~L., Chakrabarty, D., Wang, Z., \& Wachter, S.\ 2009, \apj, 700, 149 

\bibitem[Kouveliotou et al.(1994)]{kouv94} Kouveliotou, C., Fishman, G.~J., Meegan, C.~A., et al.\ 1994, \nat, 368, 125 

\bibitem[Kouveliotou et al.(2001)]{kouv01} Kouveliotou, C., Tennant, A., Woods, P.~M., et al.\ 2001, \apjl, 558, L47 

\bibitem[Kozlova et al.(2016)]{kozlova16} Kozlova, A.~V., Israel, G.~L., Svinkin, D.~S., et al.\ 2016, \mnras, 460, 2008 

\bibitem[{{Kozlova}(2016)}]{kozlova16b}
{Kozlova}, A.~V. et al. 2016, GRB Coordinates Network, 19438

\bibitem[{{Kouveliotou} {et~al.}(1998){Kouveliotou}, {Dieters}, {Strohmayer},
  {van Paradijs}, {Fishman}, {Meegan}, {Hurley}, {Kommers}, {Smith}, {Frail},
  \& {Murakami}}]{kouveliotou98}
{Kouveliotou}, C., {Dieters}, S., {Strohmayer}, T., {et~al.} 1998, \nat, 393,
  235

\bibitem[{{Kouveliotou} {et~al.}(1999){Kouveliotou}, {Strohmayer}, {Hurley},
  {van Paradijs}, {Finger}, {Dieters}, {Woods}, {Thompson}, \&
  {Duncan}}]{kouveliotou99}
{Kouveliotou}, C., {Strohmayer}, T., {Hurley}, K., {et~al.} 1999, \apjl, 510,
  L115

\bibitem[{{Lien} {et~al.}(2014){Lien}, {Barthelmy}, {Baumgartner}, {Cummings},
  {Gehrels}, {Krimm}, {Markwardt}, {Palmer}, {Sakamoto}, {Stamatikos},
  {Tueller}, \& {Ukwatta}}]{lien14a}
{Lien}, A.~Y., {Barthelmy}, S.~D., {Baumgartner}, W.~H., {et~al.} 2014, GRB
  Coordinates Network, 16522


\bibitem[{{Metzger} {et~al.}(2015){Metzger}, {Margalit}, {Kasen},
  {Quataert}}]{metzger15}
{Metzger}, B.~D., {Margalit}, B., {Kasen}, D.,  {Quataert}, E. 2015, \mnras,
  454, 3311

\bibitem[{{Olausen} \& {Kaspi}(2014)}]{magcat}
{Olausen}, S.~A., \& {Kaspi}, V.~M. 2014, \apjs, 212, 6

\bibitem[Piran et al.(2017)]{piran17} Piran, T., Nakar, E., Mazzali, P., \& Pian, E.\ 2017, arXiv:1704.08298 

\bibitem[Popov \& Prokhorov(2006)]{pop06} Popov, S.~B., \& Prokhorov, M.~E.\ 2006, \mnras, 367, 732 



\bibitem[Rea et al.(2004)]{rea04} Rea, N., Testa, V., Israel, G.~L., et al.\ 2004, \aap, 425, L5 


\bibitem[{{Rea} {et~al.}(2010){Rea}, {Esposito}, {Turolla}, {Israel}, {Zane},
  {Stella}, {Mereghetti}, {Tiengo}, {G{\"o}tz}, {G{\"o}{\u g}{\"u}{\c s}}, \&
  {Kouveliotou}}]{rea10}
{Rea}, N., {Esposito}, P., {Turolla}, R., {et~al.} 2010, Science, 330, 944

\bibitem[{{Rea} \& {Esposito}(2011)}]{rea11}
{Rea}, N., \& {Esposito}, P. 2011, Astrophysics and Space Science Proceedings,
  21, 247


\bibitem[Rea et al.(2014)]{rea14} Rea, N., Vigan{\`o}, D., Israel, G.~L., Pons, J.~A., \& Torres, D.~F.\ 2014, \apjl, 781, L17 

\bibitem[{{Rea} {et~al.}(2015){Rea}, {Gull{\'o}n}, {Pons}, {Perna}, {Dainotti},
  {Miralles}, \& {Torres}}]{rea15}
{Rea}, N., {Gull{\'o}n}, M., {Pons}, J.~A., {et~al.} 2015, \apj, 813, 92


\bibitem[Rea et al.(2016)]{rea16} Rea, N., Borghese, A., Esposito, P., et al.\ 2016, \apjl, 828, L13 


\bibitem[{{Rowlinson} {et~al.}(2013){Rowlinson}, {O'Brien}, {Metzger},
  {Tanvir}, \& {Levan}}]{rowlinson13}
{Rowlinson}, A., {O'Brien}, P.~T., {Metzger}, B.~D., {Tanvir}, N.~R., \&
  {Levan}, A.~J. 2013, \mnras, 430, 1061


\bibitem[{{Schlafly} \& {Finkbeiner}(2011)}]{schlafly}
{Schlafly}, E.~F., \& {Finkbeiner}, D.~P. 2011, \apj, 737, 103

\bibitem[Scholz et al.(2014)]{scholz14} Scholz, P., Kaspi, V.~M., \& Cumming, A.\ 2014, \apj, 786, 62 


\bibitem[Surnis et al.(2016)]{surnis16} Surnis, M., Joshi, B.~C., Maan, Y., et al.\ 2016, arXiv:1605.08276 



\bibitem[Sukhbold \& Thompson(2017)]{sukhbold17} Sukhbold, T., \& Thompson, T.~A.\ 2017, arXiv:1704.06682 

\bibitem[Tam et al.(2004)]{tam04} Tam, C.~R., Kaspi, V.~M., van Kerkwijk, M.~H., \& Durant, M.\ 2004, \apjl, 617, L53 

\bibitem[Tam et al.(2008)]{tam08} Tam, C.~R., Gavriil, F.~P., Dib, R., et al.\ 2008, \apj, 677, 503-514 


\bibitem[Tendulkar et al.(2012)]{tendulkar12} Tendulkar, S.~P., Cameron, P.~B., \& Kulkarni, S.~R.\ 2012, \apj, 761, 76 


\bibitem[Tendulkar et al.(2013)]{tendulkar13} Tendulkar, S.~P., Cameron, P.~B., \& Kulkarni, S.~R.\ 2013, \apj, 772, 31 

\bibitem[Tendulkar et al.(2017)]{tendulkar17} Tendulkar, S.~P., Kaspi, V.~M., Archibald, R.~F., \& Scholz, P.\ 2017, \apj, 841, 11 


\bibitem[{{Thompson} {et~al.}(2000){Thompson}, {Duncan}, {Woods},
  {Kouveliotou}, {Finger}, \& {van Paradijs}}]{thompson00}
{Thompson}, C., {Duncan}, R.~C., {Woods}, P.~M., {et~al.} 2000, \apj, 543, 340

\bibitem[Wachter et al.(2008)]{wachter08} Wachter, S., Ramirez-Ruiz, E., Dwarkadas, V.~V., et al.\ 2008, \nat, 453, 626 


\bibitem[Wang et al.(2006)]{wang06} Wang, Z., Chakrabarty, D., \& Kaplan, D.~L.\ 2006, \nat, 440, 772 
\bibitem[Wang et al.(2008)]{wang08} Wang, Z., Bassa, C., Kaspi, V.~M., Bryant, J.~J., \& Morrell, N.\ 2008, \apj, 679, 1443-1446 

\bibitem[{{Willingale} {et~al.}(2013){Willingale}, {Starling}, {Beardmore},
  {Tanvir}, \& {O'Brien}}]{willingale}
{Willingale}, R., {Starling}, R.~L.~C., {Beardmore}, A.~P., {Tanvir}, N.~R., \&
  {O'Brien}, P.~T. 2013, \mnras, 431, 394

\bibitem[{{Younes} {et~al.}(2015){Younes}, {Gogus}, {Kouveliotou}, \& {van der
  Hors}}]{younes15}
{Younes}, G., {Gogus}, E., {Kouveliotou}, C., \& {van der Hors}, A.~J. 2015,
  The Astronomer's Telegram, 7213
  

  \bibitem[{{Younes}(2016)}]{younes16}
{Younes}, G. 2016, GRB Coordinates Network, 19437

\bibitem[Younes et al.(2017)]{younes17} Younes, G., Kouveliotou, C., Jaodand, A., et al.\ 2017, arXiv:1702.04370 



\end{thebibliography}
\end{document}